\documentclass[aps,prl,twocolumn,superscriptaddress,showpacs,preprintnumbers,amsmath,amssymb]{revtex4-1}
\usepackage{graphicx} 
\usepackage{dcolumn}  
\usepackage{float}
\usepackage{subfigure}
\usepackage{color,soul}
\usepackage{lineno}
\usepackage[colorlinks,citecolor=blue,urlcolor=blue,bookmarks=false,hypertexnames=true]{hyperref}

\usepackage{gensymb}
\usepackage{amsmath}

\begin{document}

\preprint{\vbox{ \hbox{   }
			\hbox{Belle Preprint {\it 2020-16}}
			\hbox{KEK Preprint {\it 2020-33}}}}
			
\title{\quad\\[1.0cm] Search for lepton-number- and baryon-number-violating tau decays at Belle}

\noaffiliation
\affiliation{University of the Basque Country UPV/EHU, 48080 Bilbao}
\affiliation{University of Bonn, 53115 Bonn}
\affiliation{Brookhaven National Laboratory, Upton, New York 11973}
\affiliation{Budker Institute of Nuclear Physics SB RAS, Novosibirsk 630090}
\affiliation{Faculty of Mathematics and Physics, Charles University, 121 16 Prague}
\affiliation{Chonnam National University, Gwangju 61186}
\affiliation{University of Cincinnati, Cincinnati, Ohio 45221}
\affiliation{Deutsches Elektronen--Synchrotron, 22607 Hamburg}
\affiliation{University of Florida, Gainesville, Florida 32611}
\affiliation{Department of Physics, Fu Jen Catholic University, Taipei 24205}
\affiliation{Key Laboratory of Nuclear Physics and Ion-beam Application (MOE) and Institute of Modern Physics, Fudan University, Shanghai 200443}
\affiliation{II. Physikalisches Institut, Georg-August-Universit\"at G\"ottingen, 37073 G\"ottingen}
\affiliation{SOKENDAI (The Graduate University for Advanced Studies), Hayama 240-0193}
\affiliation{Gyeongsang National University, Jinju 52828}
\affiliation{Department of Physics and Institute of Natural Sciences, Hanyang University, Seoul 04763}
\affiliation{University of Hawaii, Honolulu, Hawaii 96822}
\affiliation{High Energy Accelerator Research Organization (KEK), Tsukuba 305-0801}
\affiliation{J-PARC Branch, KEK Theory Center, High Energy Accelerator Research Organization (KEK), Tsukuba 305-0801}
\affiliation{Higher School of Economics (HSE), Moscow 101000}
\affiliation{Forschungszentrum J\"{u}lich, 52425 J\"{u}lich}
\affiliation{IKERBASQUE, Basque Foundation for Science, 48013 Bilbao}
\affiliation{Indian Institute of Science Education and Research Mohali, SAS Nagar, 140306}
\affiliation{Indian Institute of Technology Bhubaneswar, Satya Nagar 751007}
\affiliation{Indian Institute of Technology Hyderabad, Telangana 502285}
\affiliation{Indian Institute of Technology Madras, Chennai 600036}
\affiliation{Indiana University, Bloomington, Indiana 47408}
\affiliation{Institute of High Energy Physics, Chinese Academy of Sciences, Beijing 100049}
\affiliation{Institute of High Energy Physics, Vienna 1050}
\affiliation{Institute for High Energy Physics, Protvino 142281}
\affiliation{INFN - Sezione di Napoli, 80126 Napoli}
\affiliation{INFN - Sezione di Torino, 10125 Torino}
\affiliation{Advanced Science Research Center, Japan Atomic Energy Agency, Naka 319-1195}
\affiliation{J. Stefan Institute, 1000 Ljubljana}
\affiliation{Institut f\"ur Experimentelle Teilchenphysik, Karlsruher Institut f\"ur Technologie, 76131 Karlsruhe}
\affiliation{Kavli Institute for the Physics and Mathematics of the Universe (WPI), University of Tokyo, Kashiwa 277-8583}
\affiliation{Department of Physics, Faculty of Science, King Abdulaziz University, Jeddah 21589}
\affiliation{Kitasato University, Sagamihara 252-0373}
\affiliation{Korea Institute of Science and Technology Information, Daejeon 34141}
\affiliation{Korea University, Seoul 02841}
\affiliation{Kyoto Sangyo University, Kyoto 603-8555}
\affiliation{Kyungpook National University, Daegu 41566}
\affiliation{Universit\'{e} Paris-Saclay, CNRS/IN2P3, IJCLab, 91405 Orsay}
\affiliation{P.N. Lebedev Physical Institute of the Russian Academy of Sciences, Moscow 119991}
\affiliation{Liaoning Normal University, Dalian 116029}
\affiliation{Faculty of Mathematics and Physics, University of Ljubljana, 1000 Ljubljana}
\affiliation{Ludwig Maximilians University, 80539 Munich}
\affiliation{Malaviya National Institute of Technology Jaipur, Jaipur 302017}
\affiliation{University of Maribor, 2000 Maribor}
\affiliation{Max-Planck-Institut f\"ur Physik, 80805 M\"unchen}
\affiliation{School of Physics, University of Melbourne, Victoria 3010}
\affiliation{University of Mississippi, University, Mississippi 38677}
\affiliation{University of Miyazaki, Miyazaki 889-2192}
\affiliation{Moscow Physical Engineering Institute, Moscow 115409}
\affiliation{Graduate School of Science, Nagoya University, Nagoya 464-8602}
\affiliation{Kobayashi-Maskawa Institute, Nagoya University, Nagoya 464-8602}
\affiliation{Universit\`{a} di Napoli Federico II, 80126 Napoli}
\affiliation{Nara Women's University, Nara 630-8506}
\affiliation{National Central University, Chung-li 32054}
\affiliation{National United University, Miao Li 36003}
\affiliation{Department of Physics, National Taiwan University, Taipei 10617}
\affiliation{H. Niewodniczanski Institute of Nuclear Physics, Krakow 31-342}
\affiliation{Nippon Dental University, Niigata 951-8580}
\affiliation{Niigata University, Niigata 950-2181}
\affiliation{University of Nova Gorica, 5000 Nova Gorica}
\affiliation{Novosibirsk State University, Novosibirsk 630090}
\affiliation{Osaka City University, Osaka 558-8585}
\affiliation{Pacific Northwest National Laboratory, Richland, Washington 99352}
\affiliation{Panjab University, Chandigarh 160014}
\affiliation{Peking University, Beijing 100871}
\affiliation{University of Pittsburgh, Pittsburgh, Pennsylvania 15260}
\affiliation{Punjab Agricultural University, Ludhiana 141004}
\affiliation{Research Center for Nuclear Physics, Osaka University, Osaka 567-0047}
\affiliation{Meson Science Laboratory, Cluster for Pioneering Research, RIKEN, Saitama 351-0198}
\affiliation{Department of Modern Physics and State Key Laboratory of Particle Detection and Electronics, University of Science and Technology of China, Hefei 230026}
\affiliation{Seoul National University, Seoul 08826}
\affiliation{Showa Pharmaceutical University, Tokyo 194-8543}
\affiliation{Soochow University, Suzhou 215006}
\affiliation{Soongsil University, Seoul 06978}
\affiliation{Sungkyunkwan University, Suwon 16419}
\affiliation{School of Physics, University of Sydney, New South Wales 2006}
\affiliation{Department of Physics, Faculty of Science, University of Tabuk, Tabuk 71451}
\affiliation{Tata Institute of Fundamental Research, Mumbai 400005}
\affiliation{Department of Physics, Technische Universit\"at M\"unchen, 85748 Garching}
\affiliation{School of Physics and Astronomy, Tel Aviv University, Tel Aviv 69978}
\affiliation{Toho University, Funabashi 274-8510}
\affiliation{Department of Physics, Tohoku University, Sendai 980-8578}
\affiliation{Earthquake Research Institute, University of Tokyo, Tokyo 113-0032}
\affiliation{Department of Physics, University of Tokyo, Tokyo 113-0033}
\affiliation{Tokyo Institute of Technology, Tokyo 152-8550}
\affiliation{Tokyo Metropolitan University, Tokyo 192-0397}
\affiliation{Utkal University, Bhubaneswar 751004}
\affiliation{Virginia Polytechnic Institute and State University, Blacksburg, Virginia 24061}
\affiliation{Wayne State University, Detroit, Michigan 48202}
\affiliation{Yamagata University, Yamagata 990-8560}
\affiliation{Yonsei University, Seoul 03722}
  \author{D.~Sahoo}\affiliation{Tata Institute of Fundamental Research, Mumbai 400005}\affiliation{Utkal University, Bhubaneswar 751004} 
  \author{G.~B.~Mohanty}\affiliation{Tata Institute of Fundamental Research, Mumbai 400005} 
  \author{K.~Trabelsi}\affiliation{Universit\'{e} Paris-Saclay, CNRS/IN2P3, IJCLab, 91405 Orsay} 
  \author{I.~Adachi}\affiliation{High Energy Accelerator Research Organization (KEK), Tsukuba 305-0801}\affiliation{SOKENDAI (The Graduate University for Advanced Studies), Hayama 240-0193} 
  \author{K.~Adamczyk}\affiliation{H. Niewodniczanski Institute of Nuclear Physics, Krakow 31-342} 
  \author{H.~Aihara}\affiliation{Department of Physics, University of Tokyo, Tokyo 113-0033} 
  \author{S.~Al~Said}\affiliation{Department of Physics, Faculty of Science, University of Tabuk, Tabuk 71451}\affiliation{Department of Physics, Faculty of Science, King Abdulaziz University, Jeddah 21589} 
  \author{D.~M.~Asner}\affiliation{Brookhaven National Laboratory, Upton, New York 11973} 
  \author{T.~Aushev}\affiliation{Higher School of Economics (HSE), Moscow 101000} 
  \author{R.~Ayad}\affiliation{Department of Physics, Faculty of Science, University of Tabuk, Tabuk 71451} 
% ...
\collaboration{The Belle Collaboration}

\begin{abstract}
We search for lepton-number- and baryon-number-violating decays $\tau^{-}\to\overline{p}e^{+}e^{-}$, $pe^{-}e^{-}$, $\overline{p}e^{+}\mu^{-}$, $\overline{p}e^{-}\mu^{+}$, $\overline{p}\mu^{+}\mu^{-}$, and $p\mu^{-}\mu^{-}$ using 921 fb$^{-1}$ of data, equivalent to $(841\pm12)\times 10^6$ $\tau^{+}\tau^{-}$ events, recorded with the Belle detector at the KEKB asymmetric-energy $e^{+}e^{-}$ collider.
In the absence of a signal, $90\%$ confidence-level upper limits are set on the branching fractions of these decays in the range $(1.8$--$4.0)\times 10^{-8}$.
We set the world's first limits on the first four channels and improve the existing limits by an order of magnitude for the last two channels.
\end{abstract}


\maketitle

\tighten

{\renewcommand{\thefootnote}{\fnsymbol{footnote}}}
\setcounter{footnote}{0}

As lepton flavor, lepton number and baryon number are accidental symmetries of the standard model (SM), there is no reason to expect them to be conserved in all possible particle interactions.
In fact, lepton flavor violation has already been observed in neutrino oscillations~\cite{oscl}.
While baryon number (\textit{B}) is presumed to have been violated in the early Universe, its exact mechanism still remains unknown.
To explain the matter-antimatter asymmetry observed in nature, the following three conditions, formulated by Sakharov~\cite{sakharov}, must be satisfied.
\begin{enumerate}
\item \textit{B} violation: does not yet have any experimental confirmation.
\item Violation of \textit{C} (charge conjugation) and \textit{CP} (combination of \textit{C} with parity \textit{P}): both phenomena have been observed.
\item Departure from thermal equilibrium.
\end{enumerate}
Any observation of processes involving \textit{B} violation would be a clear signal of new physics.
Such processes are studied in different scenarios of physics beyond the SM such as supersymmetry~\cite{susy}, grand unification~\cite{grnd}, and models with black holes~\cite{black}.

\textit{B} violation in charged lepton decays often implies violation of lepton number (\textit{L}).
Conservation of angular momentum in such decays would require a change of $\lvert\Delta(B-L)\rvert=0$ or $2$.
These selection rules allow for several distinct possibilities.
For $\Delta(B-L)=0$, the simplest choice is $\Delta B=\Delta L=0$, e.g., standard beta decay.
A more interesting case is $\Delta B=\Delta L=\pm 1$ obeying the $\Delta(B-L)=0$ rule, which strictly holds in the SM and is the subject of this paper.
Other intriguing possibilities are $\Delta(B-L)=2$ that include $\Delta B=-\Delta L=1$ (proton decay), $\Delta B=2$ (neutron-antineutron oscillation), and $\Delta L=2$ (neutrinoless double-beta decay).
It is important to know which one of these selection rules for $B$ or $L$ violation is chosen by nature.
This will address a profound question as to whether the violation of $B$ or $L$ individually implies the violation of $(B-L)$ as well.
If it does, it must be connected with the Majorana nature of neutrinos~\cite{mohapatra}.

We report herein a search for six \textit{L}- and \textit{B}-violating decays: $\tau^{-}\to\overline{p}e^{+}e^{-}$, $pe^{-}e^{-}$, $\overline{p}e^{+}\mu^{-}$, $\overline{p}e^{-}\mu^{+}$, $\overline{p}\mu^{+}\mu^{-}$, and $p\mu^{-}\mu^{-}$~\cite{charge} in $e^{+}e^{-}$ annihilations at Belle.
Based on $1\,{\rm fb}^{-1}$ of $pp$ collision data, LHCb~\cite{lhcb} has studied the last two channels, setting $90\%$ confidence-level (CL) upper limits on their branching fractions: ${\cal B}(\tau^{-}\to\overline{p}\mu^{+}\mu^{-})$ $< 3.3\times10^{-7}$ and ${\cal B}(\tau^{-}\to p\mu^{-}\mu^{-})$ $< 4.4\times10^{-7}$.
Using experimental bounds on proton decay, authors in Refs.~\cite{theory1,theory2,theory3} predict a branching fraction in the range of $10^{-30}$--$10^{-48}$ for these kinds of $\Delta B=\Delta L=\pm 1$ decays.

We use $711\,{\rm fb}^{-1}$ ($89\,{\rm fb}^{-1}$) of data recorded at ($60\,{\rm MeV}$ below) the $\Upsilon(4S)$ resonance with the Belle detector~\cite{belle} at the KEKB asymmetric-energy $e^{+}e^{-}$ collider~\cite{kekb}.
A sample of $121\,{\rm fb}^{-1}$ collected near the $\Upsilon(5S)$ peak is also used in this search. 

Belle is a large-solid-angle magnetic spectrometer comprising a silicon vertex detector, a 50-layer central drift chamber (CDC), an array of aerogel threshold Cherenkov counters (ACC), a barrel-like arrangement of time-of-flight scintillation counters (TOF), and a CsI(Tl) crystal electromagnetic calorimeter (ECL).
All these components are located inside a superconducting solenoid providing a magnetic field of 1.5\,T.
An iron flux return located outside the solenoid coil is instrumented with resistive plate chambers to detect $K_{L}^0$ mesons and muons (KLM).

To optimize the event selection and obtain signal detection efficiency, we use Monte Carlo (MC) simulation samples.
Signal and background events from $e^{+}e^{-}\to\tau^{+}\tau^{-}(\gamma)$ are generated by the KKMC~\cite{kkmc} program, while the subsequent decays of $\tau$ leptons are handled by TAUOLA~\cite{tauola} or PYTHIA~\cite{pyth}, and final-state radiation is included with PHOTOS~\cite{photos}.
For the signal MC samples, we generate $\tau^{+}\tau^{-}$ events, where one $\tau$ decays into $p\ell\ell^{\prime}(\ell,\ell^{\prime}=e,\mu)$, assuming a phase-space distribution, and the other $\tau$ into all SM-allowed final states (``generic decay").
Non-$\tau$ backgrounds, such as $e^{+}e^{-}\to q\overline{q}$ ($udsc$ continuum, $B\overline{B}$), Bhabha scattering, and dimuon processes are generated with EvtGen~\cite{evtgen}, BHLUMI~\cite{bhlumi}, and KKMC, respectively.
We generate two-photon mediated final states using DIAG36~\cite{aafhb} and TREPS~\cite{treps}.
The DIAG36 program is applied for the $e^{+}e^{-}q\overline{q}$ production as well as for the $e^{+}e^{-}e^{+}e^{-}$ and $e^{+}e^{-}\mu^{+}\mu^{-}$ processes.
We use TREPS to generate the $e^{+}e^{-}p\overline{p}$ final state with its cross section tuned to the known measurements.
Additionally, MC samples for suppressed decays~\cite{tauexcl} $\tau^{-}\to\pi^{-}e^{+}e^{-}\nu_{\tau}$ and  $\pi^{-}\mu^{+}\mu^{-}\nu_{\tau}$ are used to study possible background contaminations.

We follow a ``blind" analysis technique in this search, where the signal region (defined below) in data remains hidden until all of our selection criteria and background estimation methods are finalized.
Below we describe different stages of event reconstruction and selection.
All kinematic observables are measured in the laboratory frame unless stated otherwise.

At the preliminary level, we try to retain as many generic $e^{+}e^{-}\to\tau^{+}\tau^{-}$ events as possible in the sample while reducing obvious backgrounds.
Towards that end, we apply the following criteria on different kinematic variables.
Charged track and photon candidates are selected within a range of $17\degree <\theta <150\degree$, where $\theta$ is their polar angle relative to the $z$ axis (opposite the $e^{+}$ beam direction).
We require the transverse momentum ($p_{\rm T}$) of each charged track to be greater than $0.1\,{\rm GeV}$ and the energy of each photon to be greater than $0.1\,{\rm GeV}$. 
Natural units $\hbar=c=1$ are used throughout the paper.
Each track must have a distance of closest approach with respect to the interaction point (IP) within $\pm 0.5$\,cm in the transverse plane and within $\pm 3.0$\,cm along the $z$ axis.
Candidate $\tau$-pair events are required to have four charged tracks with zero net charge; this criterion greatly reduces the amount of background from high-multiplicity $e^{+}e^{-}\to q\overline{q}$ events.
We require the primary vertex, reconstructed by minimizing the sum of $\chi^{2}$'s computed with helix parameters measured for all four tracks, to be close to the IP.
Requirements on the radius, $r<1.0$\,cm, and $z$ position, $|z|<3.0$\,cm, of the event primary vertex suppress beam-related and cosmic muon backgrounds.

As two-photon mediated events contain many low-$p_{\rm T}$ tracks, a minimum threshold on the highest $p_{\rm T}$ track ($p_{\rm T}^{\rm max}>0.5\,{\rm GeV}$) provides a useful handle against such events.
This background is suppressed further by requiring either $p_{\rm T}^{\rm max}>1\,{\rm GeV}$ or $E_{\rm rec}>3\,{\rm GeV}$, where $E_{\rm rec}$ is the sum of momenta of all charged tracks and energies of all photons in the center-of-mass (CM) frame.
Additionally, we require [$E_{\rm {tot}}<9\,{\rm GeV}$, $\theta_{\rm {max}}<175\degree$, or $2<E_{{\rm ECL}}<10\,{\rm GeV}$] and [$N_{\rm {barrel}}\geq 2$, or $E_{\rm ECL}^{\rm trk}<5.3\,{\rm GeV}$], where the total energy $E_{\rm tot}=E_{\rm rec}+p_{\rm miss}^{\rm CM}$ with $p_{\rm miss}^{\rm CM}$ being the magnitude of the missing momentum in the CM frame, $\theta_{\rm max}$ is the maximum opening angle between any two tracks, $E_{\rm ECL}$ is the sum of energies deposited by all tracks and photons in the ECL, $N_{\rm barrel}$ is the number of tracks in the barrel region, given by $30\degree<\theta<130\degree$, and $E_{\rm ECL}^{\rm trk}$ is the sum of energies deposited by tracks in the ECL in the CM frame.

\begin{figure}[!h]
\begin{center}
    \includegraphics[width=0.48\textwidth]{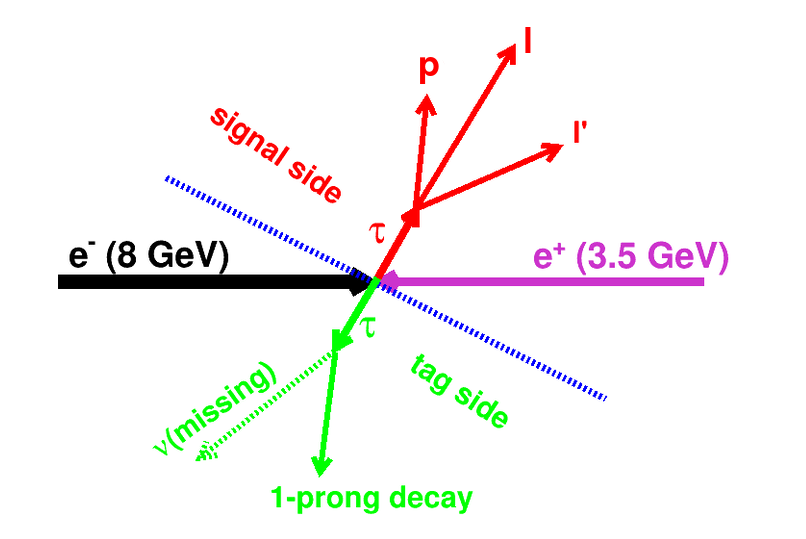}
\end{center}
\caption{A schematic of 3-1 topology defined in the CM frame. The blue dotted line divides the event into two hemispheres based on the thrust-axis direction.}
\label{fig:fig31}
\end{figure}

At the second stage of selection, we apply the following criteria to pick up candidate events that are more signal-like.
First we require the four charged tracks to be arranged in a 3-1 topology as shown in Fig.~\ref{fig:fig31}. 
This classification is done by means of the thrust axis~\cite{thrust} calculated from the observed track and photon candidates. 
One of the two hemispheres divided by the plane perpendicular to the thrust axis should contain three tracks (signal side) and the other has one track (tag side).
To reduce $e^{+}e^{-}\to q\overline{q}$ background further, we require the magnitude of the thrust to be greater than $0.9$.

As neutrinos are emitted only from the tag-side $\tau$ candidate in case of a signal, the direction of the missing momentum vector ($\vec{p}_{\rm miss}$) lies on the tag side.
The cosine of the angle between $\vec{p}_{\rm miss}$ and the momentum of the track on the tag side in the CM frame is thus required to be greater than zero.
Photons from radiative Bhabha and dimuon events are emitted in the beam direction.
Similarly, the initial-state electrons and positrons in two-photon events are emitted along the beam pipe.
To suppress these events, we require the polar angle of $\vec{p}_{\rm miss}$ to lie between $5\degree$ and $175\degree$.
The aforementioned sets of selection criteria are common to all six channels. 

We require one of the three charged tracks in the signal side to be identified as a proton or an antiproton.
It must satisfy ${\cal L}(p/K)>0.6$ and ${\cal L}(p/\pi)>0.6$, where ${\cal L}(i/j)={\cal L}_{i}/({\cal L}_{i}+{\cal L}_{j})$ with ${\cal L}_{i}$ and ${\cal L}_{j}$ being the likelihood for the track to be identified as $i$ and $j$, respectively.
The likelihood values are obtained~\cite{pidref} by combining specific ionization ($dE\!/\!dx$) measured in the CDC, the number of photoelectrons in the ACC, and the flight time from the TOF.
The proton identification efficiency with the above likelihood criteria is about $95\%$, while the probability of misidentifying a kaon or a pion as a proton is below $10\%$. 

Electrons are distinguished from charged hadrons with a likelihood ratio eID, defined as ${\cal L}_{e}/({\cal L}_{e}+{\cal L}_{\widetilde{e}})$, where ${\cal L}_{e}$ (${\cal L}_{\widetilde{e}}$) is the likelihood value for electron (not-electron) hypothesis.
These likelihoods are determined~\cite{eidref} using the ratio of the energy deposited in the ECL to the momentum measured in the CDC, the shower shape in the ECL, the matching between the position of charged-track trajectory and the cluster position in the ECL, the number of photoelectrons in the ACC, and $dE\!/\!dx$ measured in the CDC.
To recover the energy loss due to bremsstrahlung, photons are searched for in a cone of 50\,mrad around the initial direction of the electron momentum; if found, their momenta are added to that of the electron.
For muon identification an analogous likelihood ratio~\cite{muidref} is defined as $\mu$ID $={\cal  L}_{\mu}/({\cal L}_{\mu}+{\cal L}_{\pi}+{\cal L}_{K})$, where ${\cal L}_{\mu}$, ${\cal L}_{\pi}$, and ${\cal L}_{K}$ are calculated with the matching quality and penetration depth of associated hits in the KLM.
We apply eID $>0.9$ and $\mu$ID $>0.9$ to select the electron and muon candidates, respectively.
The electron (muon) identification efficiency for these criteria is $91\%$ ($85\%$) with the probability of misidentifying a pion as an electron (a muon) below $0.5\%$ ($2\%$).
The kaon-to-electron misidentification rate is negligible, while the probability of detecting a kaon as a muon is similar to that of a pion.

We apply a loose criterion ${\rm eID}<0.9$ on the $p$ or $\overline{p}$ candidate to suppress the potential misidentification of electrons as protons.
No particle identification requirement is applied for the sole track in the tag side, for which the default pion hypothesis is assumed.

The $\tau$ lepton is reconstructed by combining a proton or an antiproton with two charged lepton candidates.
A vertex fit is performed for the $\tau$ candidate reconstructed from these three charged tracks.
To identify the signal, we use two kinematic variables: the reconstructed mass $M_{\rm rec}=\sqrt{E_{p\ell\ell^{\prime}}^{2}-\vec{p}_{p\ell\ell^{\prime}}^{\,2}}$ and the energy difference $\Delta E=E^{\rm CM}_{p\ell\ell^{\prime}}-E^{\rm CM}_{\rm beam}$, where $E_{p\ell\ell^{\prime}}$ and $\vec{p}_{p\ell\ell^{\prime}}$ are the sum of energies and momenta, respectively, of the $p$, $\ell$ and $\ell^{\prime}$ candidates.
The beam energy $E^{\rm CM}_{\rm beam}$ and $E^{\rm CM}_{p\ell\ell^{\prime}}$ are calculated in the CM frame.
For signal events $M_{\rm rec}$ peaks at the nominal $\tau$ mass~\cite{pdg2020} and $\Delta E$ near zero.

\begin{figure}[!htb]
\begin{center}
 \includegraphics[width=.8\linewidth]{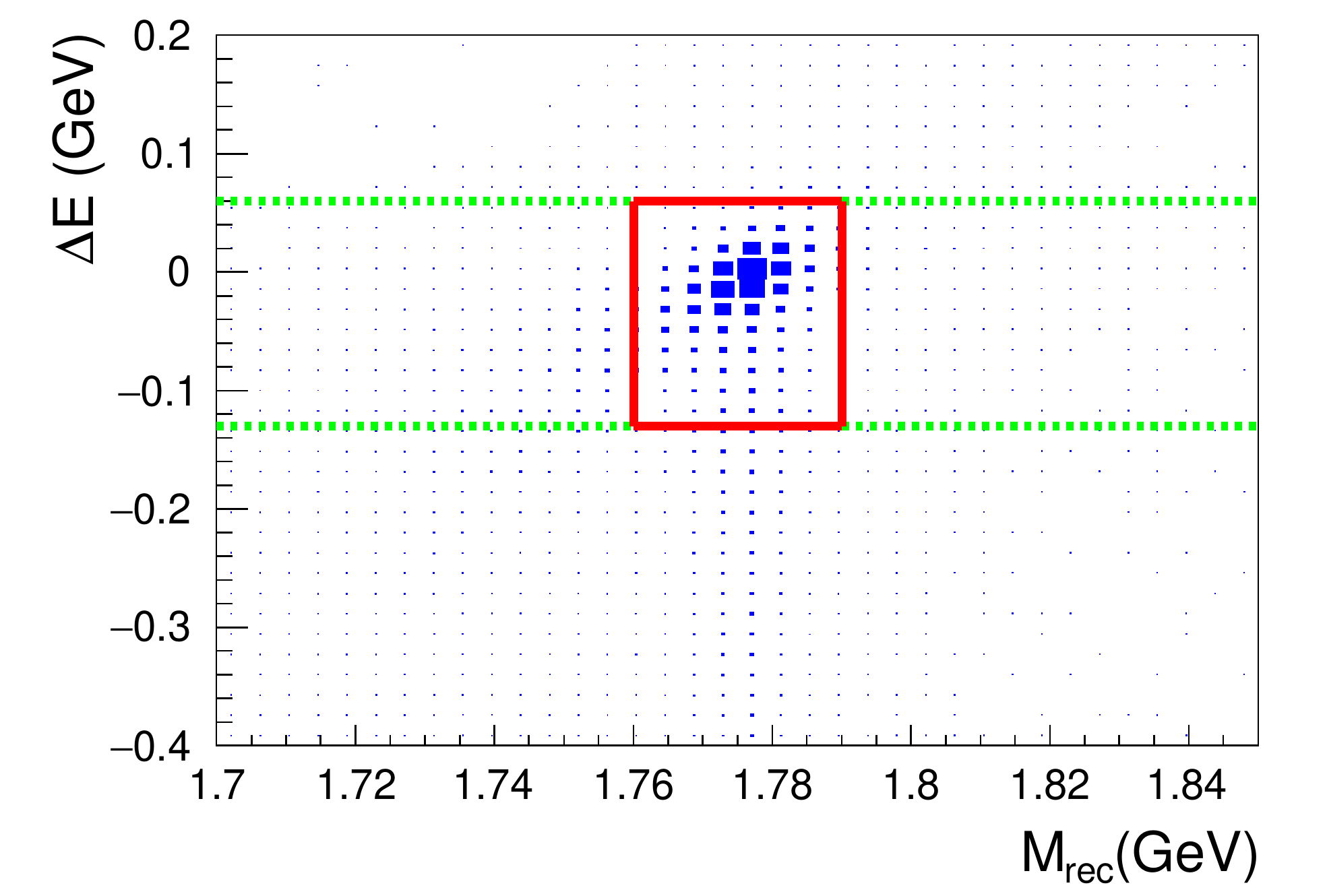}
\end{center}
\caption{$\Delta E$--$M_{\rm rec}$ distribution for the $\tau^{-}\to\overline{p}e^{+}e^{-}$ signal MC sample.
The red box denotes the signal region, the region outside it is the sideband, and the area between two green dashed lines excluding the red box is the $\Delta E$ strip.
The size of the blue filled box represents the number of events in a given bin.
For other channels these three regions are similarly defined except that the red box position is changed owing to the difference in $\Delta E$ and $M_{\rm rec}$ resolutions.}
\label{fig:chan1sig}
\end{figure}

The signal region is taken as $1.76\leq M_{\rm rec}\leq1.79\,{\rm GeV}$ and $-0.13\leq\Delta E\leq0.06\,{\rm GeV}$ for the $\tau^{-}\to\overline{p}e^{+}e^{-}$ and $\tau^{-}\to pe^{-}e^{-}$ channels (shown by the red box in Fig.~\ref{fig:chan1sig}).
Similarly, for the $\tau^{-}\to\overline{p}e^{+}\mu^{-}$ and $\tau^{-}\to\overline{p}e^{-}\mu^{+}$ channels, the signal region is defined as $1.764\leq M_{\rm rec}\leq1.789\,{\rm GeV}$ and $-0.110\leq\Delta E\leq0.055\,{\rm GeV}$.
Lastly, for the $\tau^{-}\to\overline{p}\mu^{+}\mu^{-}$ and $\tau^{-}\to p\mu^{-}\mu^{-}$ channels, the signal region is given by $1.766\leq M_{\rm rec}\leq1.787\,{\rm GeV}$ and $-0.10\leq\Delta E\leq0.05\,{\rm GeV}$.
The $M_{\rm rec}$ requirements correspond to a $\pm 3\sigma$ window and the $\Delta E$ ranges are chosen to be asymmetric [$-5\sigma,+3\sigma$] owing to the radiative tail on the negative side, where $\sigma$ is the resolution of the respective kinematic variable.
The radiative tail is the largest (smallest) for channels with two electrons (muons) in the final state.
The sideband is the $\Delta E$--$M_{\rm rec}$ region outside the signal region; we use it to check the data-MC agreement for different variables.
Similarly, the $\Delta E$ strip, indicated by the region between two green dashed lines excluding the red box in Fig.~\ref{fig:chan1sig}, is used to calculate the expected background yield in the signal region.

\begin{figure}[!htb]
\begin{center}
 \includegraphics[width=.75\linewidth]{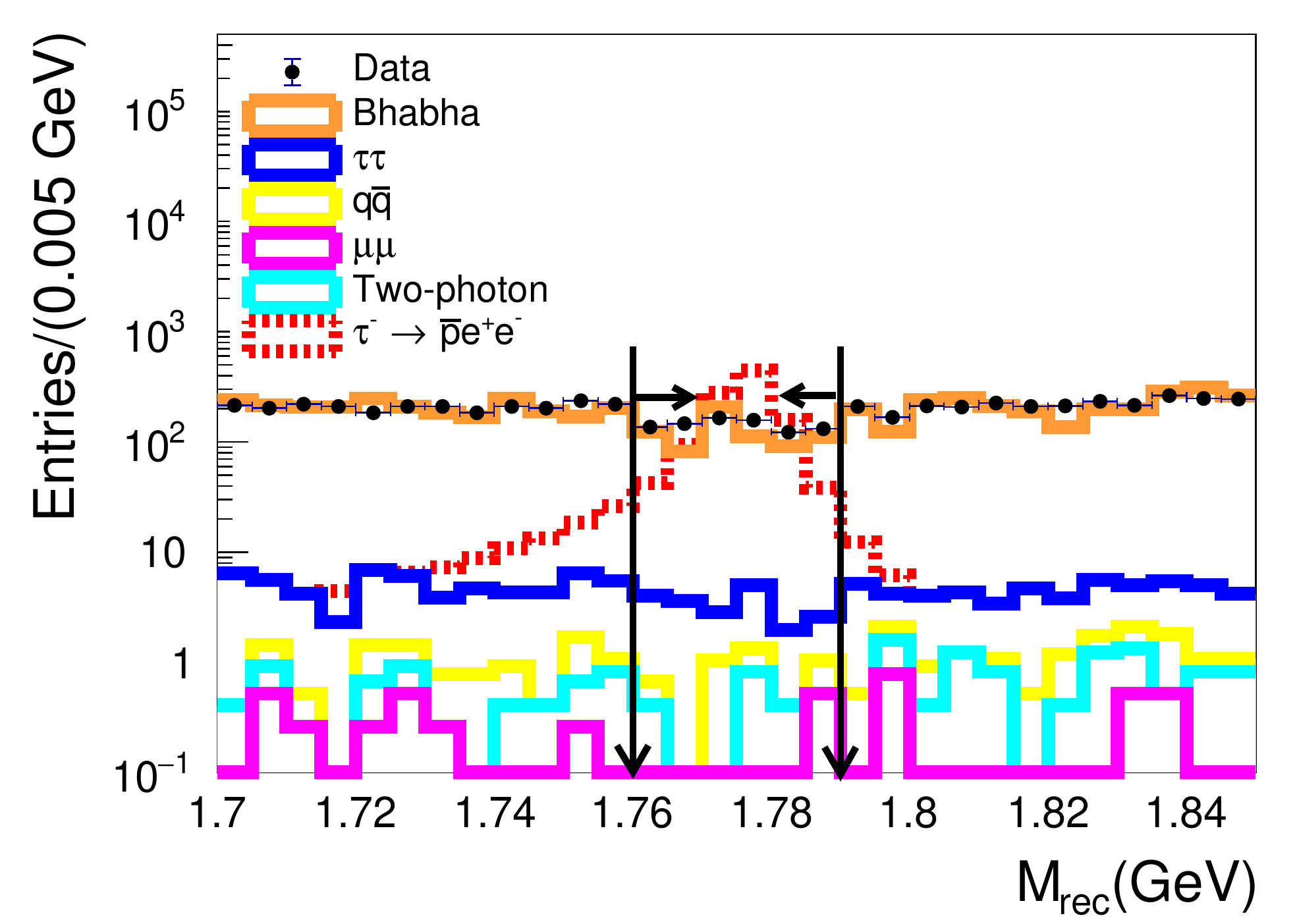}
 \includegraphics[width=.75\linewidth]{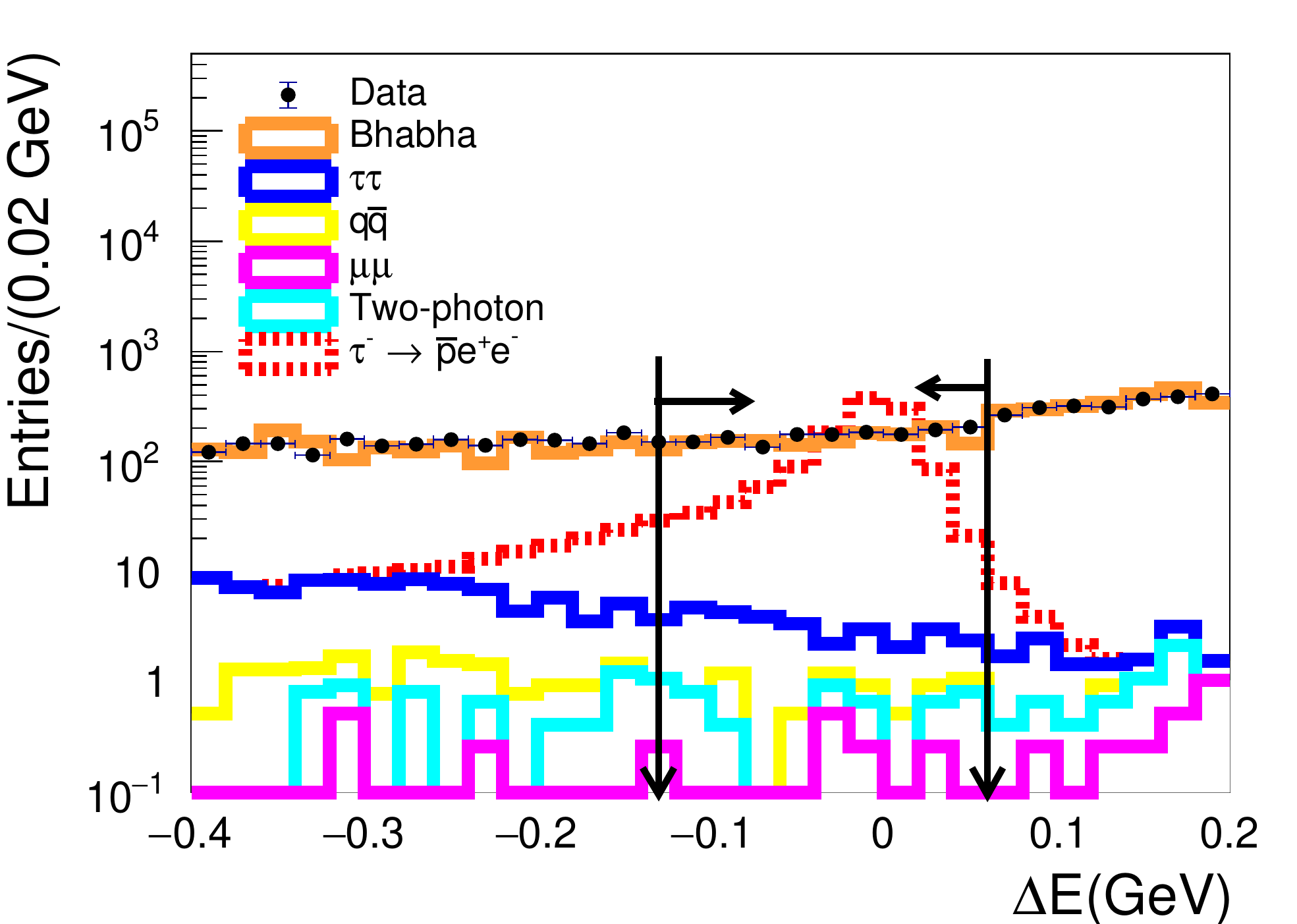}
\end{center}
\caption{$M_{\rm rec}$ and $\Delta E$ distributions in the sideband for $\tau^{-}\to\overline{p}e^{+}e^{-}$ before the photon conversion veto applied. Black arrows denote the signal region. Signal MC events are arbitrarily normalized while background MC events are scaled to the number of data events.}
\label{fig:apepemnoveto}
\end{figure}

We perform a sideband study to identify the sources of background that are dominated by events with a misidentified proton or antiproton, as well as to verify the overall data-MC agreement.
After applying the requirements used for the selection of $\tau$-pair events and charged particle identification, the $M_{\rm rec}$ and $\Delta E$ distributions for the remaining $\tau^{-}\to\overline{p}e^{+}e^{-}$ candidates in the sideband are shown in Fig.~\ref{fig:apepemnoveto}.

\begin{figure}[!h]
\begin{center}
 \includegraphics[width=.75\linewidth]{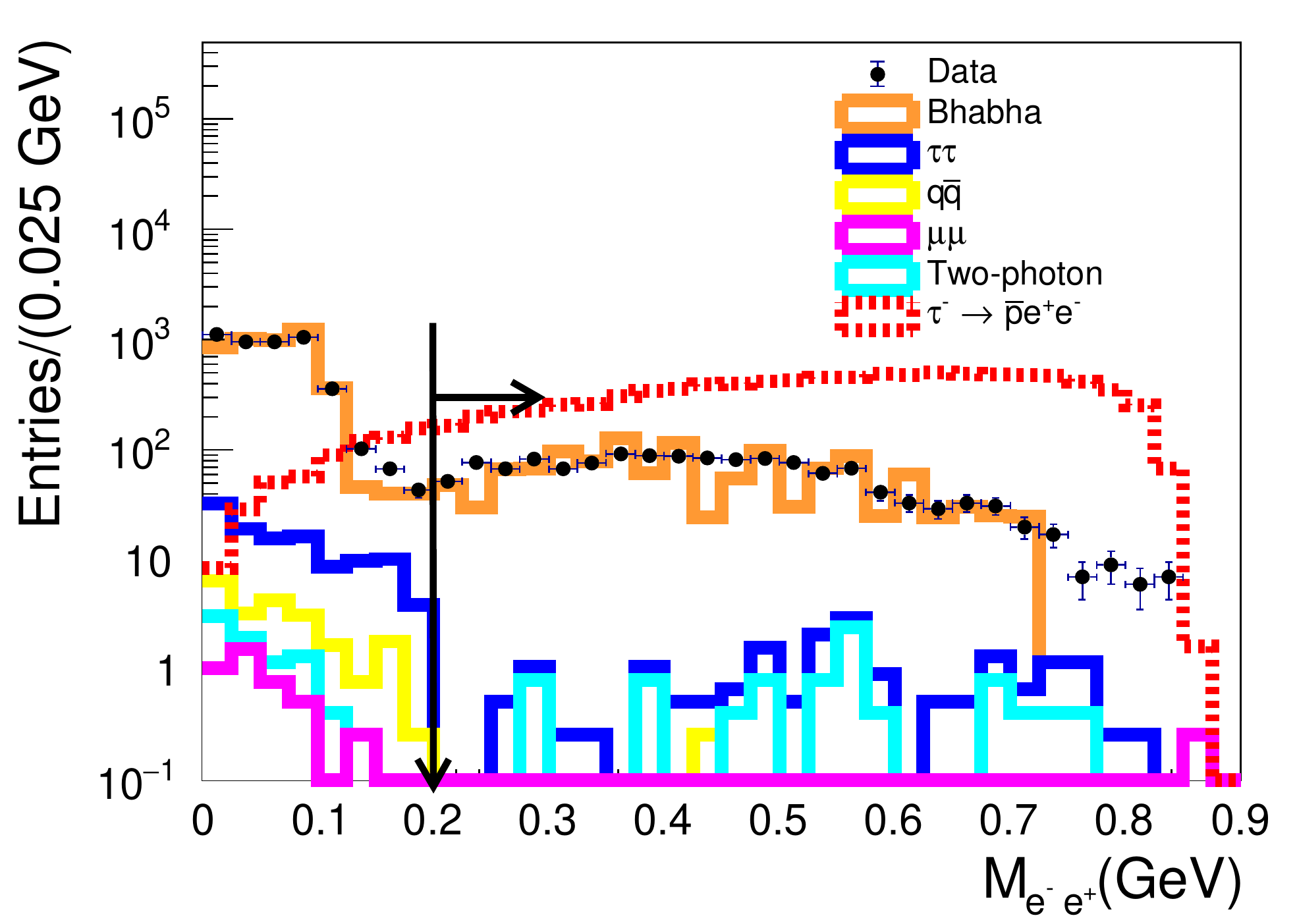}
 \includegraphics[width=.75\linewidth]{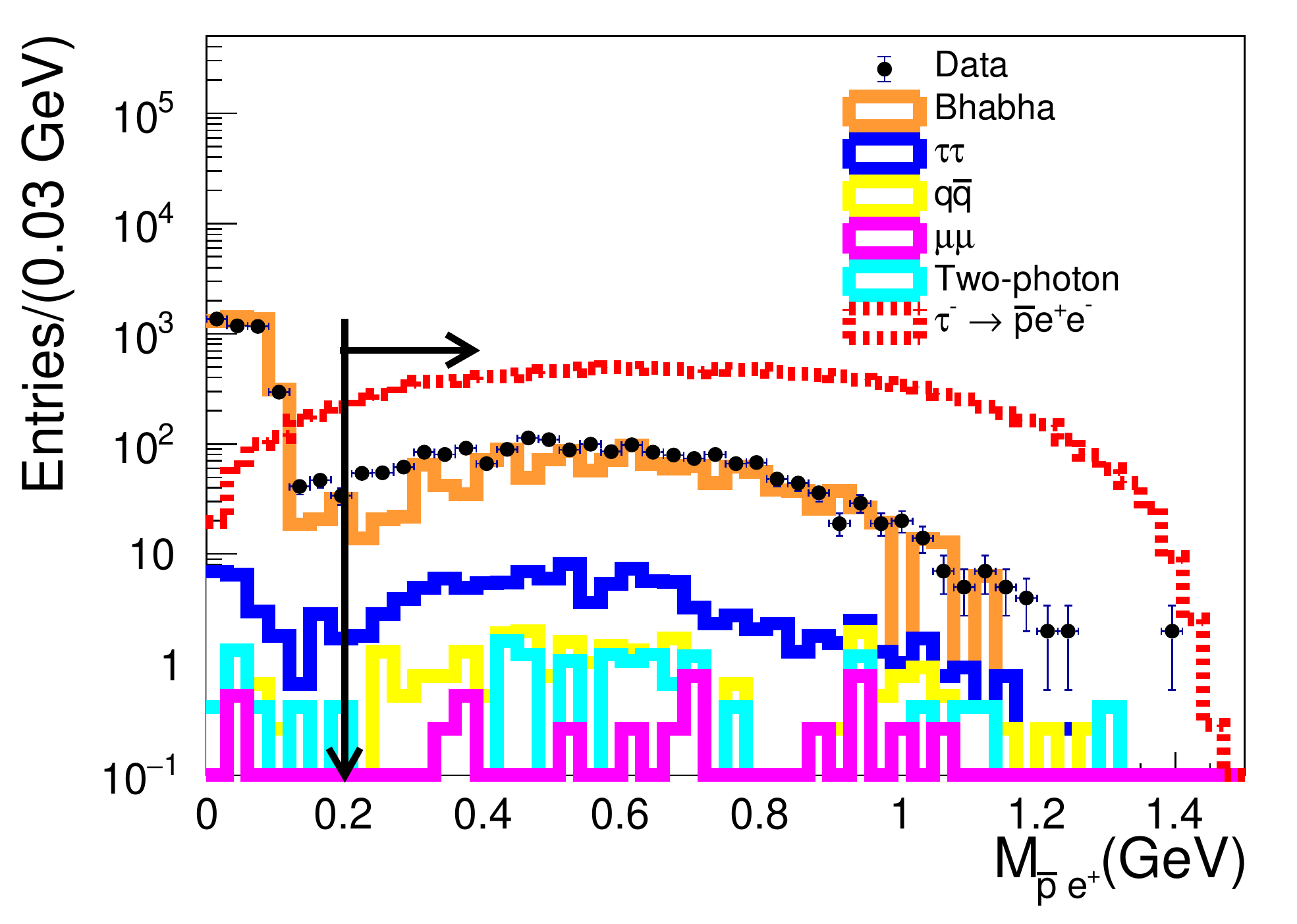}
\end{center}
\caption{$M_{e^+e^-}$ and $M_{\overline{p}e^+}$ (electron hypothesis) distributions in the $\tau^{-}\to\overline{p}e^{+}e^{-}$ sideband. Black arrows show the conversion veto position. Signal MC events are arbitrarily normalized while background MC events are scaled to the number of data events.}
\label{fig:conversion}
\end{figure}

Photon conversion in the detector material constitutes a major background for the $\tau^{-}\to\overline{p}e^{+}e^{-}$ channel.
To suppress it, we require the invariant mass of two oppositely charged track pairs $M_{e^{+}e^{-}}$ and $M_{\overline{p}e^{+}}$, calculated under the electron hypothesis,  to be greater than $0.2\,{\rm GeV}$ (Fig.~\ref{fig:conversion}).
The remaining contribution is largely from radiative Bhabha events leading to the final state of $e^{+}e^{-}e^{+}e^{-}$.
As there are four electrons in the final state, a maximum threshold of $10\,{\rm GeV}$ on the sum of their ECL cluster energies helps suppress these backgrounds.
We apply the same set of criteria for $\tau^{-}\to pe^{-}e^{-}$.

In the $\tau^{-}\to\overline{p}e^{+}\mu^{-}$ channel, the presence of $\overline{p}$ and $e^{+}$ in the final state leads to a possible background from photon conversion.
A conversion veto ($M_{\overline{p}e^{+}} >0.2\,{\rm GeV}$) as described above is applied to suppress its contamination; here the electron hypothesis is assumed for the antiproton track.
We apply no conversion veto for $\tau^{-}\to\overline{p}e^{-}\mu^{+}$ in absence of a peak in $M_{\overline{p}\mu^{+}}$.

We check the possibility of electrons from photon conversion faking muons in $\tau^{-}\to p\mu^{-}\mu^{-}$. This arises from radiative dimuon events, where one of the electrons from $\gamma\to e^{+}e^{-}$ is misidentified as a proton and the other as a muon.
For the latter to happen, the electron must pick up some KLM hits of the signal-side muon while both have the same charge.
On calculating the invariant mass of the proton and muon tracks under the electron hypothesis, we find a small peak and apply the veto $M_{p\mu^{-}}>0.2\,{\rm GeV}$ to suppress the conversion. 
As both muons have the opposite charge in $\tau^{-}\to\overline{p}\mu^{+}\mu^{-}$, there is no chance for an electron to fake a muon.
Indeed, a negligible peaking contribution is found in the $M_{\overline{p}\mu^{+}}$ distribution, requiring no conversion veto.

From the MC study the following sources of backgrounds remain after the final selection.
We find contributions mainly from $\tau$ decays, two-photon, and $q\overline{q}$ events for $\tau^{-}\to\overline{p}e^{+}e^{-}$; and $\tau$ decay and two-photon events for $\tau^{-}\to pe^{-}e^{-}$.
Similarly, $\tau$ decays, dimuon, and $q\overline{q}$ events are the residual contributors for $\tau^{-}\to\overline{p}e^{+}\mu^{-}$; and  $\tau$ decays, dimuon, $q\overline{q}$, and two-photon events for $\tau^{-}\to\overline{p}e^{-}\mu^{+}$.
For $\tau^{-}\to p\mu^{-}\mu^{-}$ and $\tau^{-}\to\overline{p}\mu^{+}\mu^{-}$ we have contributions mostly from $\tau$ decays and $q\overline{q}$ events.
The backgrounds listed above for a given channel are in the descending order of their contributions.
While calculating the background contribution from $\tau$ decays, we use the exclusive MC samples for suppressed decays, where appropriate.

To calculate the background in the signal region, we assume a uniform background distribution along the $M_{\rm rec}$ axis in Fig.~\ref{fig:chan1sig}.
The assumption is validated with MC samples before applying the method to data.
As only a few events survive our final set of selections, it becomes a challenge to know the background shape in the $M_{\rm rec}$--$\Delta E$ plane.
Instead of changing our selections channel-by-channel, we release the proton identification requirement for all six channels to check the background shape in the sideband.
While this alleviates the issue of low event yields, we find for $\tau^{-}\to p\mu^{-}\mu^{-}$ and $\overline{p}\mu^{+}\mu^{-}$ the negative $\Delta E$ region is overpopulated, mostly owing to $\pi\to\mu$ misidentification in generic $\tau$ decays.
Similarly, in case of $\tau^{-}\to\overline{p}e^{+}e^{-}$ and $\overline{p}e^{-}\mu^{+}$ the positive $\Delta E$ region has a higher event yield coming from two-photon and radiative dimuon events.
On the other hand, for all the channels the $\Delta E$ strip is found to have a uniform event density in $M_{\rm rec}$.
Therefore, we calculate the background yield in the signal region based on the number of events found in the $ \Delta E$ strip in lieu of the full sideband.
The expected numbers of background events in the signal region with uncertainties are listed in Table~{\ref{tab:tab1}} for all channels.

\begin{figure}[!htb]
\begin{center}

  \includegraphics[width=0.5\textwidth]{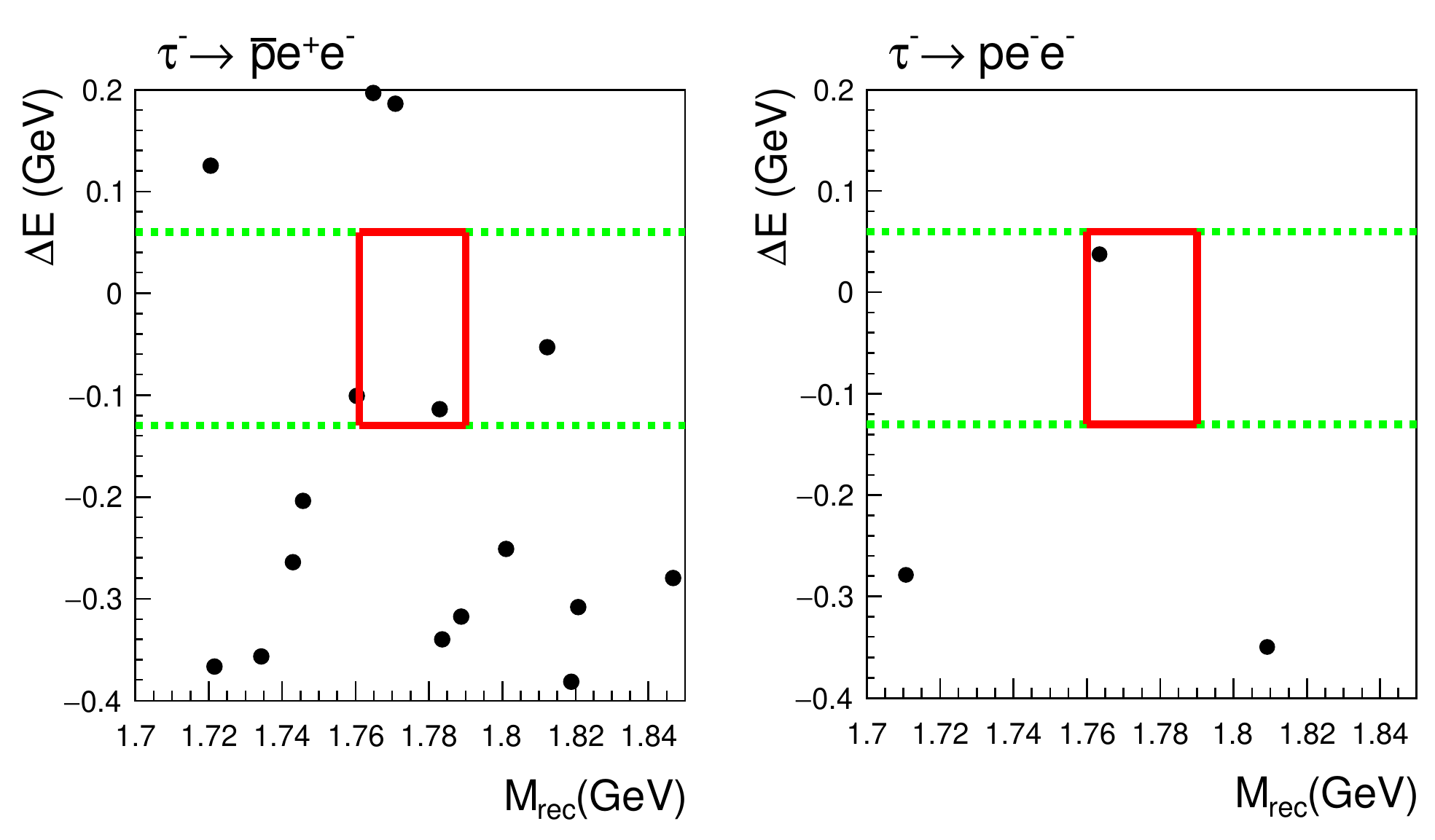}
 \includegraphics[width=0.5\textwidth]{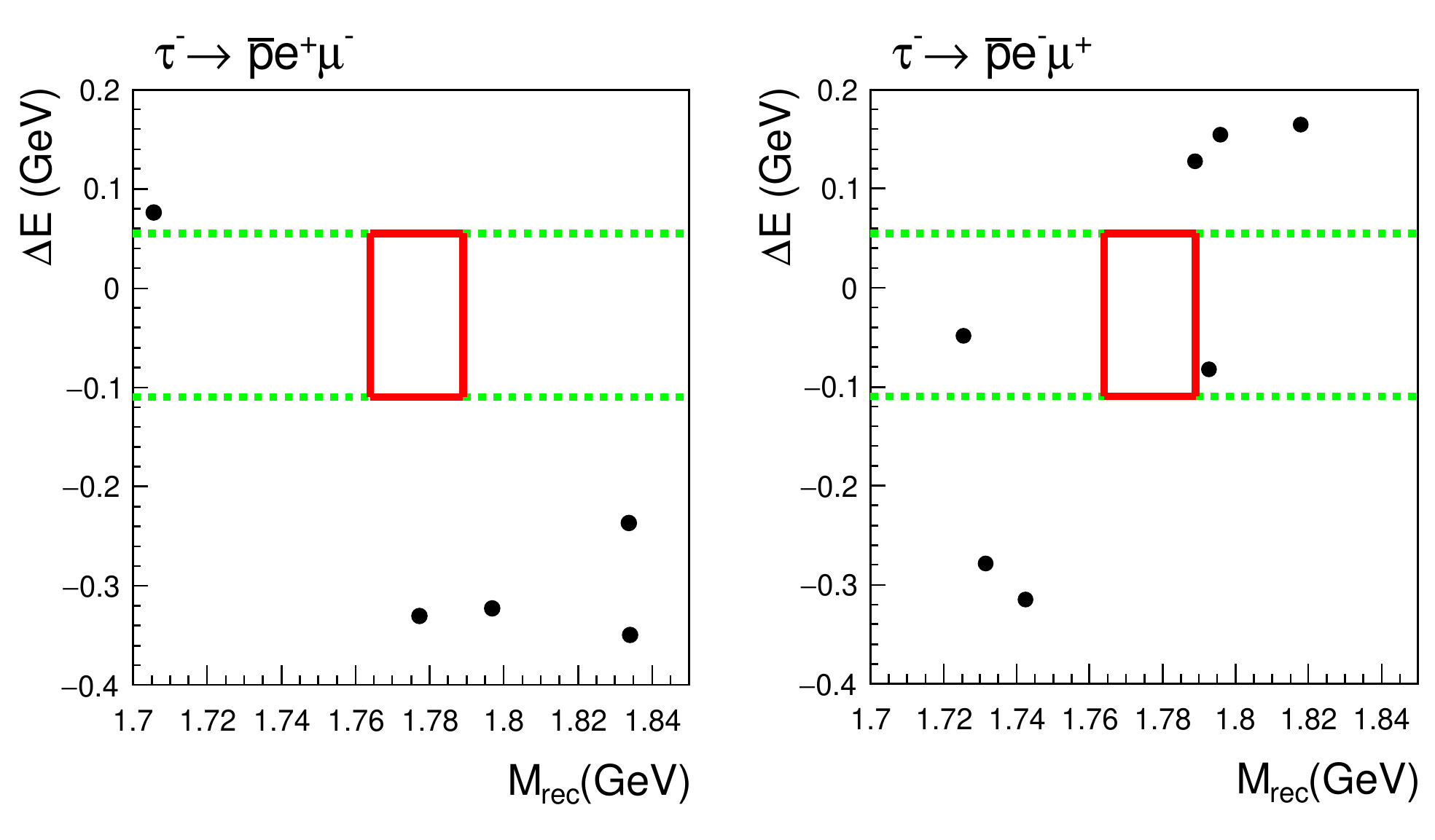}
 \includegraphics[width=0.5\textwidth]{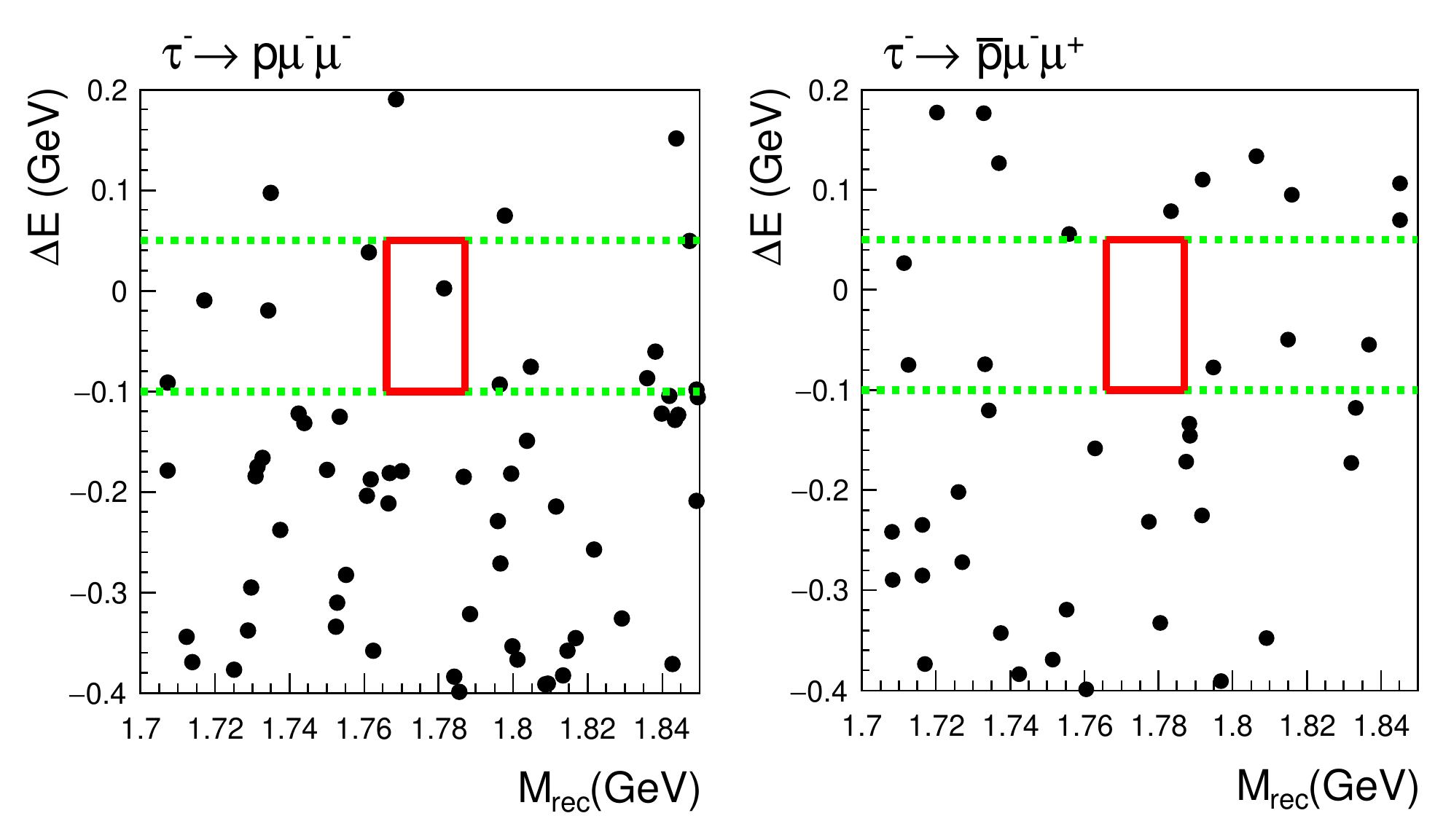}
\end{center}
\caption{$\Delta E$--$M_{\rm rec}$ distributions where the red box denotes the signal region and the green $\Delta E$ strip is used to calculate the expected background. Black dots represent the data.}
\label{fig:fig1}
\end{figure}

For $\tau^{-}\to pe^{-}e^{-}$ and $\overline{p}e^{+}\mu^{-}$ channels, no events survive in the $\Delta E$ strip as shown in Fig.~\ref{fig:fig1}.
In these two cases, we use the following method to get an approximate background yield in the strip. 
As the $\tau^{-}\to p\mu^{-}\mu^{-}$ channel has the most number of events, we take the ratio of events in its lower sideband with and without applying proton identification.
We multiply this ratio by the number of events found in $\tau^{-}\to pe^{-}e^{-}$ and $\overline{p}e^{+}\mu^{-}$ without proton identification requirement to get an approximate background yield in the $\Delta E$ strip, from which the expected number of background in the signal region is calculated.
We have checked that this method gives a background yield consistent with that directly obtained from the $\Delta E$ strip for other four channels.

We calculate the systematic uncertainties arising from various sources.
The uncertainties due to lepton identification are $2.3\%$ per electron and $2.0\%$ per muon.
Similarly, the proton identification uncertainty is $0.5\%$.
Tracking efficiency uncertainty is $0.35\%$ per track, totaling $1.4\%$ for four tracks in the final state.
For the systematic uncertainty due to efficiency variation, we take half of the maximum spread in efficiency with respect to its average value found in the invariant-mass variables: $M_{p\ell}$, $M_{p\ell^{\prime}}$, and $M_{\ell\ell^{\prime}}$.
The uncertainty in the trigger efficiency studied with a dedicated trigger simulation program is found to be $1.2\%$~\cite{tauexcl}.
All these multiplicative contributions are added in quadrature to get a total systematic uncertainty in efficiency.
The uncertainty associated with integrated luminosity is $1.4\%$, and that due to the $e^{+}e^{-}\to\tau^{+}\tau^{-}$ cross section is $0.3\%$.
Both contribute as an uncertainty to the number of $\tau$ pairs used in the upper limit calculation (see below). 

There is one event observed in data in each of the $\tau^{-}\to\overline{p}e^{+}e^{-}$, $pe^{-}e^{-}$, and $p\mu^{-}\mu^{-}$ channels as shown in Fig.~\ref{fig:fig1}.
We find no events in the signal region in the case of $\tau^{-}\to\overline{p}e^{-}\mu^{+}$, $\overline{p}e^{+}\mu^{-}$, and $\overline{p}\mu^{-}\mu^{+}$. 
As the number of events observed in the signal region is consistent with the background prediction, we calculate an upper limit using the Feldman-Cousins method~\cite{fc-method}.
The $90\%$ CL upper limit on the signal yield ($N_{\rm sig}^{\rm UL}$) is obtained with the POLE program~\cite{pole} based on the number of observed data and expected background events, the uncertainty in background, as well as uncertainties in efficiency and number of $\tau$ pairs. 
The upper limit on the branching fraction is then:
\begin{equation}
{\cal B}(\tau^{-}\to p\mu^{-}\mu^{-})<\dfrac{N_{\rm sig}^{\rm UL}}{2N_{\tau\tau}\epsilon},
\end{equation}
where the detection efficiency in the signal region ($\epsilon$) is determined by multiplying the off-line selection efficiency by the trigger efficiency, and $N_{\tau\tau}=\sigma_{\tau\tau}{\cal L}_{\rm int}=(841\pm12)\times 10^{6}$ is the number of $\tau$ pairs expected in $921\,{\rm fb}^{-1}$ of data.
The trigger efficiency is about $90\%$ for all the channels. 
In Table~\ref{tab:tab1} we list results for all channels. 
The obtained upper limits range from $1.8\times 10^{-8}$ to $4.0\times 10^{-8}$.

\begin{table}[!h]
\caption{Signal detection efficiency, number of expected background events ($N _{\rm bkg}$), number of observed data events ($N _{\rm obs}$), 90$\%$ CL upper limits on the signal yield and branching fraction for various decay channels.}
\label{tab:tab1}
\begin{center}
\begin{tabular}{cccccc}
\hline\hline
Channel&$\epsilon\,(\%)$&$N _{\rm bkg} $&$N_{\rm obs}$&$N_{\rm sig}^{\rm UL}$&${\cal B}\,(\times 10^{-8})$\\
\hline
$\tau^{-}\to\overline{p}e^{+}e^{-}$     & $7.8$ & $0.50\pm 0.35$ & $1$ & $3.9$ & $<3.0$\\
$\tau^{-}\to pe^{-}e^{-}$               & $8.0$ & $0.23\pm 0.07$ & $1$ & $4.1$ & $<3.0$\\
$\tau^{-}\to\overline{p}e^{+}\mu^{-}$   & $6.5$ & $0.22\pm 0.06$ & $0$ & $2.2$ & $<2.0$\\
$\tau^{-}\to\overline{p}e^{-}\mu^{+}$   & $6.9$ & $0.40\pm 0.28$ & $0$ & $2.1$ & $<1.8$\\
$\tau^{-}\to p\mu^{-}\mu^{-}$           & $4.6$ & $1.30\pm 0.46$ & $1$ & $3.1$ & $<4.0$\\
$\tau^{-}\to\overline{p}\mu^{-}\mu^{+}$ & $5.0$ & $1.14\pm 0.43$ & $0$ & $1.5$ & $<1.8$\\
\hline
\end{tabular}
\end{center}
\end{table}

In summary, we have searched for six lepton-number- and baryon-number-violating $ \tau$ decays into a proton or an antiproton and two charged leptons using $921\,{\rm fb^{-1}}$ of data.
In the case of $\tau^{-}\to p\mu^{-}\mu^{-}$ and $\overline{p}\mu^{-}\mu^{+}$, our limits are improved by an order of magnitude compared to LHCb~\cite{lhcb}.
For the remaining four channels, we set limits for the first time.
These results would be useful in the current and future pursuits of baryon number violation.

We acknowledge fruitful discussions with and helpful suggestions from S. Mahapatra (Utkal University), E. Passemar (Indiana University), and P.~S. Bhupal Dev (Washington University).
We thank the KEKB group for excellent operation of the accelerator; the KEK cryogenics group for efficient solenoid operations; and the KEK computer group, the NII, and 
PNNL/EMSL for valuable computing and SINET5 network support.
We acknowledge support from MEXT, JSPS and Nagoya's TLPRC (Japan); ARC (Australia); FWF (Austria); NSFC and CCEPP (China); MSMT (Czechia); CZF, DFG, EXC153, and VS (Germany); DAE and DST (India); INFN (Italy); MOE, MSIP, NRF, RSRI, FLRFAS project, GSDC of KISTI and KREONET/GLORIAD (Korea); MNiSW and NCN (Poland); MSHE, Agreement 14.W03.31.0026 (Russia); University of Tabuk (Saudi Arabia); ARRS (Slovenia); IKERBASQUE (Spain); SNSF (Switzerland); MOE and MOST (Taiwan); and DOE and NSF (USA).


\begin{thebibliography}{99}

\bibitem{oscl} Y. Fukuda {\it et al.} (Super-Kamiokande Collaboration), Phys. Rev. Lett. {\bf 81}, 1562 (1998); Q.~R. Ahmad {\it et al.} (SNO Collaboration), Phys. Rev. Lett. {\bf 89}, 011301 (2002).
\bibitem{sakharov} A.~D. Sakharov, JETP Lett. {\bf 5}, 24 (1967).
\bibitem{susy} S.~P. Martin, A supersymmetry primer, in Perspectives
on Supersymmetry (World Scientific, Singapore, 1998),
pp. 1–98.
\bibitem{grnd} H. Georgi and S.~L. Glashow, Phys. Rev. Lett. {\bf 32}, 438 (1974).
\bibitem{black} J.~D. Bekenstein, Phys. Rev. D {\bf 5}, 1239 (1972).
\bibitem{mohapatra} R.~N. Mohapatra and R.~E. Marshak, Phys. Rev. Lett. {\bf 44}, 1316 (1980); Erratum: Phys. Rev. Lett. {\bf 44}, 1644 (1980).
\bibitem{charge} Inclusion of charge-conjugate processes is implied unless explicitly stated otherwise.
\bibitem{lhcb} R. Aaij {\it et al.} (LHCb Collaboration), Phys. Lett. B {\bf 724}, 36 (2013).
\bibitem{theory1} W.~J. Marciano, Nucl. Phys. B, Proc. Suppl. {\bf 40}, 3 (1995).
\bibitem{theory2} W.-S. Hou, M. Nagashima, and A. Soddu, Phys. Rev. D {\bf 72}, 095001 (2005).
\bibitem{theory3} J. Fuentes-Martin, J. Portoles, and P. Ruiz-Femenia, JHEP {\bf 1501}, 134 (2015).
\bibitem{belle} A. Abashian {\it et al.} (Belle Collaboration), Nucl. Instrum. Methods Phys. Res., Sec. A {\bf 479}, 117 (2002); also see Section 2 in J. Brodzicka {\it et al.}, Prog. Theor. Exp. Phys. {\bf 2012}, 04D001 (2012).
\bibitem{kekb} S. Kurokawa and E. Kikutani, Nucl. Instrum. Methods Phys. Res., Sec. A {\bf 499}, 1 (2003), and other papers included in this volume; T. Abe {\it et al.}, Prog. Theor. Exp. Phys. {\bf 2013}, 03A001 (2013) and following articles up to 03A011.
\bibitem{kkmc} S. Jadach, B. Ward, and Z. W\c{a}s, Comp. Phys. Commun. {\bf 130}, 260 (2000).
\bibitem{tauola} S. Jadach, Z. W\c{a}s, R. Decker, and J.~H. K\"{u}hn, Comp. Phys. Commun. {\bf 76}, 361 (1993).
\bibitem{pyth} T. Sj\"{o}strand, S. Mrenna, and P. Skands, JHEP {\bf 0605}, 026 (2006).
\bibitem{photos} E. Barberio and Z. W\c{a}s, Comput. Phys. Commun. {\bf 79}, 291 (1994).
\bibitem{evtgen} D.~J. Lange, Nucl. Instrum. Methods Phys. Res., Sect. A {\bf 462}, 152 (2001).
\bibitem{bhlumi} S. Jadach, E. Richter-W\c{a}s, B.~F.~L. Ward, and Z. W\c{a}s, Comp. Phys. Commun. {\bf 70}, 305 (1992).
\bibitem{aafhb} F.~A. Berends, M. Daverveldt, and R.~H. Kleiss, Comp. Phys. Commun. {\bf 40}, 285 (1986).
\bibitem{treps} S. Uehara, arXiv:1310.0157v1 (2013).
\bibitem{tauexcl} Y. Jin {\it et al.} (Belle Collaboration), Phys. Rev. D {\bf 100}, 071101(R) (2019).
\bibitem{thrust} S. Brandt, C. Peyrou, R. Sosnowski, and A. Wroblewski, Phys. Lett. {\bf 12}, 57 (1964); E. Farhi, Phys. Rev. Lett. {\bf 39}, 1587 (1977).
\bibitem{pidref} E. Nakano, Nucl. Instrum. Methods Phys. Res., Sect. A {\bf 494}, 402 (2002).
\bibitem{eidref} K. Hanagaki, H. Kakuno, H. Ikeda, T. Iijima, and T. Tsukamoto, Nucl. Instrum. Methods Phys. Res., Sect. A {\bf 485}, 490 (2002).
\bibitem{muidref} A. Abashian {\it et al.}, Nucl. Instrum. Methods Phys. Res., Sect. A {\bf 491}, 69 (2002).
\bibitem{pdg2020} P.~A. Zyla {\it et al.} (Particle Data Group), Prog. Theor. Exp. Phys. {\bf 2020}, 083C01 (2020).
\bibitem{fc-method} G.~J. Feldman and R.~D. Cousins, Phys. Rev. D {\bf 57}, 3873 (1998).
\bibitem{pole} J. Conrad, O. Botner, A. Hallgren, and C. Pérez de los Heros, Phys. Rev. D {\bf 67}, 012002 (2003); also see \url{https://github.com/ftegenfe/polepp} (updated version of the POLE program).
\end{thebibliography}
\end{document}